\documentstyle[twoside,fleqn,espcrc2,epsf]{article}

\def\captionfont{\normalsize} 

\def\captionskip{\vspace{-9mm}}        
\def\gnuplotcaptionskip{\vspace{-0mm}} 

\newcommand{\ewxy}[2]{\setlength{\epsfxsize}{#2}\epsfbox[10 60 640 570]{#1}}

\def\seq {\! = \!}        

\let\ga=\gamma

\let\De=\Delta

\let\del=\nabla
\let\si=\sigma

\let\om=\omega
\let\Om=\Omega

\let\<=\langle
\let\>=\rangle

\let\txt=\textstyle

\def\eqn#1{(\ref{#1})}  

\def\beq{\begin{equation}}
\def\eeq{\end{equation}}
\def\ba{\begin{array}}
\def\bea{\begin{eqnarray}}
\def\ea{\end{array}}
\def\eea{\end{eqnarray}}

\def\o{\over}
\def\no{\noindent}

\def\slash{\!\!\!\!/\,}

\def\nl{\hfil\break}

\def\comment#1{ \hbox{[{\it Comment suppressed here.}\/]} }
\def\hide#1{}
\def\O{ {\cal O} }

\def\eg{{\it e.g.}}
\def\comment#1{ }

\def\IR{\relax{\rm I\kern-.18em R}}

\def\st{\rule[-1.5ex]{0em}{4ex}} 

\newcommand{\skipover}[1]{}

\def\half {{\txt {1\over 2}}}

\def\psib{{\bar\psi}}
\def\Omb{{\bar\Om}}

\def\={\!=\!}
\def\+{\,+\,}
\def\-{\,-\,}

\def\delc{{\del_{\! c \,}}}
\def\delslc{{\del\slash}_{\! c \,}}

\long\def \omit #1 {}

\newsavebox{\eqlabel}

\catcode`\@=11
\newlength{\numblen}
\newsavebox{\eqnumb}
\def\@eqnnum{%
\savebox{\eqnumb}{\rm (\theequation)}%
\settowidth{\numblen}{\usebox{\eqnumb}}%
\makebox[\numblen][l]{\usebox{\eqnumb}~~~\usebox{\eqlabel}}%
}
\catcode`\@=12

\newenvironment{equationwithlabel}[1]{ %
  \savebox{\eqlabel}{#1}
  \begin{equation}\label{#1} }{\end{equation}\savebox{\eqlabel}{~}}

\newcommand{\beql}[1]{\begin{equationwithlabel}{#1}}
\newcommand{\eeql}{\end{equationwithlabel}}
\catcode`\@=11
\def\@versim#1#2{\lower 2\p@\vbox{\baselineskip\z@skip\lineskip+1\p@
    \ialign{$\m@th#1\hfil##\hfil$\crcr#2\crcr\sim\crcr}}}
\def\gapp{\mathrel{\mathpalette\@versim>}}
\def\lapp{\mathrel{\mathpalette\@versim<}}

\def \reset@font {}

\catcode`\@=12

\newcommand{\AmS}{{\protect\the\textfont2
  A\kern-.1667em\lower.5ex\hbox{M}\kern-.125emS}}

\title{Towards Highly Improved Quark Actions
\thanks{Based on talks by M.G.A. and T.R.K.
Research conducted at the Cornell Theory Center and at SCRI, supported by
DOE grants DE-FG05-85ER25000, DE-FG05-92ER40742,
DE-FG02-90ER40542 and the Monell Foundation.}}

\author{M.G. Alford
  \address{School of Natural Sciences,
  Institute for Advanced Study, Princeton, NJ 08540 },
    T.R. Klassen
  \address{SCRI, Florida State University, Tallahassee, FL 32306},
   G.P. Lepage
   \address{Newman Laboratory of Nuclear Studies,
   Cornell University, Ithaca, NY 14853,  USA }
        }

\begin{document}

\begin{abstract}
%
%

We describe two ideas useful in the construction of highly improved
quark actions for
simulations on coarse lattices: (i) Field transformations solve
the doubler problem without destroying tree-level improvement for
on- or off-shell quantities.
The simplest example is the Sheikholeslami-Wohlert (clover) action.
Going to the next order of improvement
yields   the class of D234 actions.
(ii) {\it Anisotropic} lattices with $a_t < a_s$ are useful because they
push up the energy of unphysical branches of the dispersion relation
(which are generic to highly improved actions),
allow accurate mass determinations for particles
with bad signal/noise properties (glueballs, P-state mesons), and enable one
to simulate heavy quarks within a relativistic framework.
{}~ We present first simulation results for the quenched light hadron and
charmonium spectra obtained with a D234 action on anisotropic lattices.
\end{abstract}

\kern -10ex
\maketitle

\setlength{\arraycolsep}{0.1em}

\section{Introduction}\label{sec:intro}

The last few years have seen a revival of the Symanzik improvement
program~\cite{Sym,LW,SW}.  The impetus was largely provided by tadpole
improvement (TI)~\cite{TI}, which yields a simple first estimate of
the large perturbative corrections to the tree-level coefficients in an
improved action. In the case of non-relativistic
QCD~\cite{NRQCD} and pure glue~\cite{Glue}, the TI
of tree-level coefficients has lead to accurate
results on lattices as coarse as $a\seq0.4$~fm.

Improved glue has classical $O(a^4)$ errors, and after TI any
remaining quantum errors (of which the leading $\O(a^2 \alpha)$ piece
is known~\cite{LW} and can be eliminated) seem to be
small~\cite{Glue}.  For quarks, the Sheikholeslami-Wohlert (SW)
action~\cite{SW}, which has classical $\O(a^2)$ errors, is the most
improved action that has been used in large-scale simulations.  For
full QCD simulations to be possible on coarse lattices, it therefore
seems mandatory to design more highly improved quark
actions~\cite{oldD234,FW}.  Since exploratory simulations will have
to be performed in the quenched approximation, for which the
$a\seq 0$  hadron spectrum is not  a priori  known,
it is important to have simple {\it improvement tests} for a quark
action. As such we propose:\nl
(i) The ``effective velocity of
light'' $c({\bf p})$ (defined by
$c({\bf p})^2 = (E({\bf p})^2 - E(0)^2)/|{\bf p}|^2$) of various hadrons.
Its deviation from 1 for
non-zero masses and momenta is a measure of scaling violations. \nl
(ii) The ``$r$-test''.
For actions with Wilson and clover terms one can obtain a rough
estimate of the correctness of the coefficient 
of the clover term, 
by seeing whether the spectrum is invariant under (small) changes of the
Wilson parameter $r$. \nl
%
(iii) The scaling of dimensionless quantities such as
$m_N/m_\rho$, $m_\De/m_\rho$, and $J$~\cite{J}. \nl
(iv) The lattice spacing dependence of
 the rho mass in units of
one of the standard scales (string tension, charmonium
S--P splitting, etc).  This is a sensitive probe of scaling violations,
but it requires that the
scaling, systematic and measurement
errors of the
chosen scale be considerably smaller then those of the rho mass.

Last year we presented results for the ``isotropic D234'' quark
action~\cite{oldD234}, which has classical $\O(a^3)$ errors.
It gave very impressive results for all hadron
mass ratios and the dispersion relation of mesons,
much better than results from the SW action. However, the absolute value of
the rho mass was almost identical to that of the SW action~\cite{SCRI} for the
three lattice spacings where they could be compared.
To understand this somewhat incongruous behavior
we have started to explore a variety of improved quark actions.
Our aim is to disentangle the various sources of scaling errors. 

One possible source of scaling violations is the unphysical branches
in the quark dispersion relation that are generically present in
actions improved beyond $\O(a)$. This has lead us to the study of
actions on anisotropic lattices, and, in particular, to devise a simple
procedure to construct actions that are tree-level improved to
any order and doubler-free on an arbitrary lattice~\cite{HIQA}.
Working on anisotropic lattices with $a_t < a_s$  has several advantages: \nl
(a) By decreasing $a_t$ one can decouple the unphysical branches by pushing
up their energy. \nl
(b) Larger and more easily identifiable effective mass plateaux lead to more
accurate and confident mass determinations. This is important for particles
with bad signal/noise properties, like P-state mesons and
especially~\cite{MorPea}
glueballs.\nl
(c) Small $a_t$ allows one to simulate heavy quarks within a relativistic
framework without the prohibitive cost of a fine spatial lattice.

The disadvantage of anisotropic lattices is that there are more
independent coefficients in an improved  action, and
they have to be tuned to restore space--time exchange symmetry.
One hopes that suitable TI gives reasonable estimates of (most of) the
coefficients, also on anisotropic lattices,
and our tests indicate that in general (\cite{oldD234,aniso}, Sect.~3)
this appears to be the case.
It is always better, of course,
to tune them non-perturbatively. 
We briefly discuss this possibility in sect.~4.
Tuning the clover term is important already for the isotropic case. It could
solve the one problem of the isotropic D234 action, its low rho
mass~\cite{oldD234}.

\section{Classically Improved Quark Actions}\label{sec:actions}

\no
There is a simple procedure to construct
tree-level improved
quark actions on an arbitrary lattice without a doubler problem.
Let $a_\mu$ be the
lattice spacings of a generic hypercubic lattice. [$a_0\equiv a_t$;
when the spatial
lattice spacings $a_i$ are identical they will be denoted by $a_s$,
such a lattice with $\xi \equiv a_s/a_t$ will be referred to as a
`$\xi$:1 lattice']. The  idea is described in four steps:
\vskip 0.5mm

\no
(i) Start with a naive (improved) fermion action,
$\psib_c M_c \psi_c$, $M_c = \delslc + m_c$, where $\delc_\mu$ is an (improved)
discretization of the continuum Dirac operator $D_\mu$, differing from it
at $\O(a_\mu^n)$, say.
The subscript `c' stands for  `continuum-like', since this action is
manifestly improved to $\O(a^n)$. However, this action will
have doublers.
The simplest choice of  $\delc_\mu$, which leads to the SW action, is the
standard anti-hermitean covariant derivative,
\beq
 \del_\mu \psi(x) \equiv {1\o 2 a_\mu} \bigl[\psi_{+\mu} - \psi_{-\mu} \bigr]
            = D_\mu + \O(a_\mu^2)
\eeq
with $\psi_{\pm\mu} \equiv U_{\pm\mu}(x)\psi(x\pm\mu)$.
We will be interested in higher orders of improvement,
as exemplified by the improved derivative
\beq
\delc_\mu \, = \, \del_\mu - {1\over 6} \, a^2_\mu \del_\mu\Delta_\mu
          \, = \,  D_\mu + \O(a_\mu^4) \, .
\eeq
Here $\Delta_\mu$ is the standard second order
derivative,
\beq
\De_\mu \psi(x)  \, \equiv \, {1\o a_\mu^2}
                \bigl[ \psi_{+\mu} + \psi_{-\mu} -2\psi(x) \bigr] \, .
\eeq

\no
(ii) To cast the action in a form where doublers can easily be eliminated,
perform a field transformation $\psi_c  = \Om \psi, \psib_c = \psib \Omb$,
so that $\psib_c M_c \psi_c = \psib M_\Om \psi$,
where $ M_\Om \equiv \Omb M_c \Om$.
We choose $\Om = \Omb$ (when acting to the right) and
\beq\label{prodcvop}
 \Omb ~\Om = 1 - {1\over 2} r a_0 ~(\delslc - m_c) + \O(a^n) ~,
\eeq
in terms of the (initially) free parameter $r$, the Wilson parameter.
The transformed fermion operator $M_\Om$  reads at this point
\beq\label{MOm}
 M_\Om = m_c (1 + \half r a_0 m_c) + \delslc
       - {1\over 2} r a_0 \del\slash^2_c + \O(a^n) \, .
\eeq
This action still has doublers.

\no
(iii) To remove the doublers, use
\beq\label{delcslsq}
\delslc^2 \, = \, \De_\mu - {1\over 12} a_\mu^2 \De_\mu^2 +
  {1\over 2}
  \si\! \cdot \! F + \ldots
\eeq
to the appropriate order to eliminate $\del_\mu$ in terms of $\De_\mu$.
Here $\si\! \cdot \! F \equiv \sum_{\mu\nu} \si_{\mu\nu} F_{\mu\nu}$
is the clover term, with $F_{\mu\nu}$ a lattice representation of the field
strength, improved to the appropriate order.
 Since in momentum space $\Delta_\mu$ does not vanish at
the edge of the Brillouin zone, the action obtained by the above truncation
is doubler-free. This action, denoted by $\psib M \psi$, is on-shell
improved up to $\O(a^n)$ errors. If one is only interested in spectral
quantities, one can stop here.

\no
(iv) To (classically) also improve non-spectral quantities, {\it undo}
the field transformation, which, \eg, amounts to using the propagator
$G=\Om M^{-1} \Omb$~\cite{Heat}.
Note that this step does not reintroduce the doublers,
since it was the truncation in step~(iii), not the field
transformation {\it per se}, that eliminated them.

We emphasize that in this approach one does {\it not} have to separately
check the improvement of the interactions; it follows from the fact that
they are improved for the naive action we started with, and this fact is
not affected by the field transformation
(the Jacobian of the latter can be ignored to the order we are working).

Let us explicitly write down the actions $M$ obtained
for $n=2$ and 4:
The SW action
\beq\label{MSW}
 M_{{\rm SW}} =
  m_0 + \del\slash - {1\over 2} r a_0
 \biggl( \sum_{\mu} \Delta_\mu  +\half \si \! \cdot \! F \biggr) \,
,
\eeq
and              the D234 action
\beq\label{MD234}
\ba{rcl}
 M_{{\rm D234}} &=& m_0 +
\sum_\mu \ga_\mu \del_\mu ( 1 - b_\mu a_\mu^2 \De_\mu ) \nonumber \\[1.5mm]
 &&  +c_\mu a_\mu^3 \De_\mu^2
 \, - {1\over 2} r a_0
 \bigl( \sum_\mu \Delta_\mu + \half \si \! \cdot \! F \bigr) \, ,
\ea
\eeq
where $m_0 \equiv m_c (1 + {1\over 2} r a_0 m_c)$ in both cases, and
\beq\label{D234n}
b_\mu = 1/6,\qquad c_\mu = r a_0 /(24 a_\mu) \, .
\eeq

We now choose $r$ to lead to as few and high lying unphysical branches
in the free dispersion relation as possible.  The general D234
action~\eqn{MD234} has four branches $E=E({\bf p})$ (not counting the
particle anti-particle symmetry $E \! \leftrightarrow \! -E$), of which three
are unphysical. One easily shows:\nl
$\bullet$
For $b_0 \seq 2 c_0$ there will be at most two unphysical branches.\nl
$\bullet$
If furthermore   $r = 1 - 2 b_0$ or  $b_0 = 0$ there is (at most) one
unphysical branch.\nl
$\bullet$
There are no unphysical branches if and only if
  $r\seq 1, b_0 \seq c_0 \seq 0$.

For the SW action $r=1$ is therefore the canonical choice.
For a D234 action obtained as above one will have at
least one unphysical branch; exactly one if one chooses
$r={2\o 3}$ and $c_0 = {1\o 12}$, in addition to $b_0={1\o 6}$.
To obtain an isotropic D234 action (on an isotropic lattice) one must choose
all   $b_\mu={1\o 6}$, $c_\mu={1\o 12}$ and $r={2\o 3}$.
Since the $c_\mu$ violate~\eqn{D234n}, this action~\cite{oldD234} has $\O(a^3)$
errors.
Like the $r=1$ SW action, it can be coded very efficiently using
the ``projection trick''.

\begin{figure}[t]
\vskip 11mm          
\ewxy{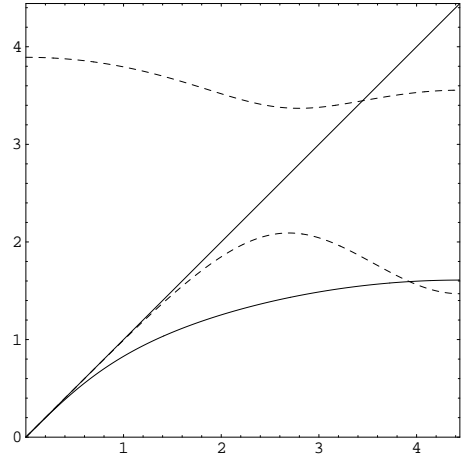}{80mm}
\captionskip
\caption{{\protect \captionfont    
 $a_s E({\bf p})$ vs $a_s |{\bf p}|$, with
${\bf p} \propto (1,1,0)$, for massless SW action on a 1:1 lattice (solid),
and D234(${2\over 3}$) on a 2:1 lattice (dashed). Continuum fermions
(thin solid) are shown for comparison.
}}
\label{fig:WilDiiiimzero}
\end{figure}

\begin{figure}[t]
\vskip 8mm   
\ewxy{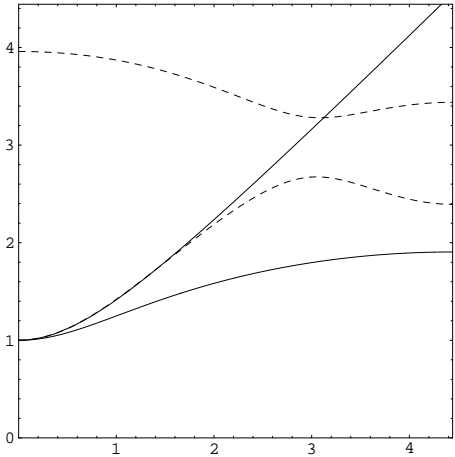}{80mm}
\captionskip
\caption{{\protect \captionfont
As in figure~1, for masses such that
$a_s E(0)=1$ in all cases.
}}
\label{fig:WilDiiiimone}
\end{figure}

Requiring just one unphysical branch is not compatible with~\eqn{D234n}.
Suitably modifying $c_0$ (and/or $b_0$ at $\O(a_0)$) to have just one
branch leads to actions with $\O(a_0^3,a_\mu^4)$ errors.
On anisotropic lattices with $a_s/a_t \geq 2$, say, one can presumably
tolerate $\O(a_0^3)$ errors. In the {\it free} case it is even possible to
avoid these $\O(a_0^3)$ errors; this is achieved~\cite{HIQA}
by a more complicated field
transformation than~\eqn{prodcvop}. It leads to a D234 action where
(some of) the coefficients have a slight mass dependence.
Its spatial coefficients satisfy~\eqn{D234n} and for the temporal ones,
$r=1-2b_0=1-4c_0={2\over 3} (1+{3\over 4} a_0 m_c )/( 1+ {7\over 12} a_0 m_c)$.
We will denote this action as
D234(${2\over 3}$), labelling it by the value of $r$
at $m_c=0$.
This is the action used in our preliminary simulations.
For details and some variations of the above procedure we refer to~\cite{HIQA}.

In figures~\ref{fig:WilDiiiimzero} and~\ref{fig:WilDiiiimone} we compare the
free quark dispersion relation of D234(${2\o 3}$) on a 2:1 with that of SW on a
1:1 lattice. Note particularly the (expected) dramatic improvement
observed
 in the heavy mass case.

\section{Simulation Results}

\def\figc{
\begin{figure}[htbp]
\vskip -3mm
\setlength{\unitlength}{0.240900pt}
\ifx\plotpoint\undefined\newsavebox{\plotpoint}\fi
\sbox{\plotpoint}{\rule[-0.200pt]{0.400pt}{0.400pt}}%
\begin{picture}(944,793)(0,0)
\font\gnuplot=cmr10 at 10pt
\gnuplot
\sbox{\plotpoint}{\rule[-0.200pt]{0.400pt}{0.400pt}}%
\put(220.0,113.0){\rule[-0.200pt]{0.400pt}{158.271pt}}
\put(220.0,137.0){\rule[-0.200pt]{4.818pt}{0.400pt}}
\put(198,137){\makebox(0,0)[r]{$0.6$}}
\put(860.0,137.0){\rule[-0.200pt]{4.818pt}{0.400pt}}
\put(220.0,259.0){\rule[-0.200pt]{4.818pt}{0.400pt}}
\put(198,259){\makebox(0,0)[r]{$0.7$}}
\put(860.0,259.0){\rule[-0.200pt]{4.818pt}{0.400pt}}
\put(220.0,381.0){\rule[-0.200pt]{4.818pt}{0.400pt}}
\put(198,381){\makebox(0,0)[r]{$0.8$}}
\put(860.0,381.0){\rule[-0.200pt]{4.818pt}{0.400pt}}
\put(220.0,502.0){\rule[-0.200pt]{4.818pt}{0.400pt}}
\put(198,502){\makebox(0,0)[r]{$0.9$}}
\put(860.0,502.0){\rule[-0.200pt]{4.818pt}{0.400pt}}
\put(220.0,624.0){\rule[-0.200pt]{4.818pt}{0.400pt}}
\put(198,624){\makebox(0,0)[r]{$1.0$}}
\put(860.0,624.0){\rule[-0.200pt]{4.818pt}{0.400pt}}
\put(220.0,746.0){\rule[-0.200pt]{4.818pt}{0.400pt}}
\put(198,746){\makebox(0,0)[r]{$1.1$}}
\put(860.0,746.0){\rule[-0.200pt]{4.818pt}{0.400pt}}
\put(220.0,113.0){\rule[-0.200pt]{0.400pt}{4.818pt}}
\put(220,68){\makebox(0,0){$0$}}
\put(220.0,750.0){\rule[-0.200pt]{0.400pt}{4.818pt}}
\put(322.0,113.0){\rule[-0.200pt]{0.400pt}{4.818pt}}
\put(322,68){\makebox(0,0){$0.2$}}
\put(322.0,750.0){\rule[-0.200pt]{0.400pt}{4.818pt}}
\put(423.0,113.0){\rule[-0.200pt]{0.400pt}{4.818pt}}
\put(423,68){\makebox(0,0){$0.4$}}
\put(423.0,750.0){\rule[-0.200pt]{0.400pt}{4.818pt}}
\put(525.0,113.0){\rule[-0.200pt]{0.400pt}{4.818pt}}
\put(525,68){\makebox(0,0){$0.6$}}
\put(525.0,750.0){\rule[-0.200pt]{0.400pt}{4.818pt}}
\put(626.0,113.0){\rule[-0.200pt]{0.400pt}{4.818pt}}
\put(626,68){\makebox(0,0){$0.8$}}
\put(626.0,750.0){\rule[-0.200pt]{0.400pt}{4.818pt}}
\put(728.0,113.0){\rule[-0.200pt]{0.400pt}{4.818pt}}
\put(728,68){\makebox(0,0){$1.0$}}
\put(728.0,750.0){\rule[-0.200pt]{0.400pt}{4.818pt}}
\put(829.0,113.0){\rule[-0.200pt]{0.400pt}{4.818pt}}
\put(829,68){\makebox(0,0){$1.2$}}
\put(829.0,750.0){\rule[-0.200pt]{0.400pt}{4.818pt}}
\put(220.0,113.0){\rule[-0.200pt]{158.994pt}{0.400pt}}
\put(880.0,113.0){\rule[-0.200pt]{0.400pt}{158.271pt}}
\put(220.0,770.0){\rule[-0.200pt]{158.994pt}{0.400pt}}
\put(67,441){\makebox(0,0){{\normalsize $c({\bf p})$}}}
\put(550,23){\makebox(0,0){\lower 6mm\hbox{{\normalsize $|{\bf p}|/{\rm
GeV}$}}}}
\put(220.0,113.0){\rule[-0.200pt]{0.400pt}{158.271pt}}
\put(509,641){\circle*{12}}
\put(629,670){\circle*{12}}
\put(720,691){\circle*{12}}
\put(798,620){\circle*{12}}
\put(509.0,629.0){\rule[-0.200pt]{0.400pt}{5.782pt}}
\put(499.0,629.0){\rule[-0.200pt]{4.818pt}{0.400pt}}
\put(499.0,653.0){\rule[-0.200pt]{4.818pt}{0.400pt}}
\put(629.0,648.0){\rule[-0.200pt]{0.400pt}{10.600pt}}
\put(619.0,648.0){\rule[-0.200pt]{4.818pt}{0.400pt}}
\put(619.0,692.0){\rule[-0.200pt]{4.818pt}{0.400pt}}
\put(720.0,659.0){\rule[-0.200pt]{0.400pt}{15.418pt}}
\put(710.0,659.0){\rule[-0.200pt]{4.818pt}{0.400pt}}
\put(710.0,723.0){\rule[-0.200pt]{4.818pt}{0.400pt}}
\put(798.0,581.0){\rule[-0.200pt]{0.400pt}{18.790pt}}
\put(788.0,581.0){\rule[-0.200pt]{4.818pt}{0.400pt}}
\put(788.0,659.0){\rule[-0.200pt]{4.818pt}{0.400pt}}
\put(509,617){\circle{12}}
\put(629,618){\circle{12}}
\put(509.0,602.0){\rule[-0.200pt]{0.400pt}{6.986pt}}
\put(499.0,602.0){\rule[-0.200pt]{4.818pt}{0.400pt}}
\put(499.0,631.0){\rule[-0.200pt]{4.818pt}{0.400pt}}
\put(629.0,598.0){\rule[-0.200pt]{0.400pt}{9.395pt}}
\put(619.0,598.0){\rule[-0.200pt]{4.818pt}{0.400pt}}
\put(619.0,637.0){\rule[-0.200pt]{4.818pt}{0.400pt}}
\put(542,451){\raisebox{-.8pt}{\makebox(0,0){$\Diamond$}}}
\put(676,459){\raisebox{-.8pt}{\makebox(0,0){$\Diamond$}}}
\put(542.0,431.0){\rule[-0.200pt]{0.400pt}{9.877pt}}
\put(532.0,431.0){\rule[-0.200pt]{4.818pt}{0.400pt}}
\put(532.0,472.0){\rule[-0.200pt]{4.818pt}{0.400pt}}
\put(676.0,437.0){\rule[-0.200pt]{0.400pt}{10.359pt}}
\put(666.0,437.0){\rule[-0.200pt]{4.818pt}{0.400pt}}
\put(666.0,480.0){\rule[-0.200pt]{4.818pt}{0.400pt}}
\put(542,308){\makebox(0,0){$\times$}}
\put(676,314){\makebox(0,0){$\times$}}
\put(542.0,276.0){\rule[-0.200pt]{0.400pt}{15.177pt}}
\put(532.0,276.0){\rule[-0.200pt]{4.818pt}{0.400pt}}
\put(532.0,339.0){\rule[-0.200pt]{4.818pt}{0.400pt}}
\put(676.0,297.0){\rule[-0.200pt]{0.400pt}{8.191pt}}
\put(666.0,297.0){\rule[-0.200pt]{4.818pt}{0.400pt}}
\put(666.0,331.0){\rule[-0.200pt]{4.818pt}{0.400pt}}
\sbox{\plotpoint}{\rule[-0.500pt]{1.000pt}{1.000pt}}%
\put(220,624){\usebox{\plotpoint}}
\put(220.00,624.00){\usebox{\plotpoint}}
\multiput(227,624)(20.756,0.000){0}{\usebox{\plotpoint}}
\multiput(233,624)(20.756,0.000){0}{\usebox{\plotpoint}}
\put(240.76,624.00){\usebox{\plotpoint}}
\multiput(247,624)(20.756,0.000){0}{\usebox{\plotpoint}}
\multiput(253,624)(20.756,0.000){0}{\usebox{\plotpoint}}
\put(261.51,624.00){\usebox{\plotpoint}}
\multiput(267,624)(20.756,0.000){0}{\usebox{\plotpoint}}
\multiput(273,624)(20.756,0.000){0}{\usebox{\plotpoint}}
\put(282.27,624.00){\usebox{\plotpoint}}
\multiput(287,624)(20.756,0.000){0}{\usebox{\plotpoint}}
\multiput(293,624)(20.756,0.000){0}{\usebox{\plotpoint}}
\put(303.02,624.00){\usebox{\plotpoint}}
\multiput(307,624)(20.756,0.000){0}{\usebox{\plotpoint}}
\multiput(313,624)(20.756,0.000){0}{\usebox{\plotpoint}}
\put(323.78,624.00){\usebox{\plotpoint}}
\multiput(327,624)(20.756,0.000){0}{\usebox{\plotpoint}}
\multiput(333,624)(20.756,0.000){0}{\usebox{\plotpoint}}
\put(344.53,624.00){\usebox{\plotpoint}}
\multiput(347,624)(20.756,0.000){0}{\usebox{\plotpoint}}
\multiput(353,624)(20.756,0.000){0}{\usebox{\plotpoint}}
\put(365.29,624.00){\usebox{\plotpoint}}
\multiput(367,624)(20.756,0.000){0}{\usebox{\plotpoint}}
\multiput(373,624)(20.756,0.000){0}{\usebox{\plotpoint}}
\put(386.04,624.00){\usebox{\plotpoint}}
\multiput(387,624)(20.756,0.000){0}{\usebox{\plotpoint}}
\multiput(393,624)(20.756,0.000){0}{\usebox{\plotpoint}}
\put(406.80,624.00){\usebox{\plotpoint}}
\multiput(407,624)(20.756,0.000){0}{\usebox{\plotpoint}}
\multiput(413,624)(20.756,0.000){0}{\usebox{\plotpoint}}
\multiput(420,624)(20.756,0.000){0}{\usebox{\plotpoint}}
\put(427.56,624.00){\usebox{\plotpoint}}
\multiput(433,624)(20.756,0.000){0}{\usebox{\plotpoint}}
\multiput(440,624)(20.756,0.000){0}{\usebox{\plotpoint}}
\put(448.31,624.00){\usebox{\plotpoint}}
\multiput(453,624)(20.756,0.000){0}{\usebox{\plotpoint}}
\multiput(460,624)(20.756,0.000){0}{\usebox{\plotpoint}}
\put(469.07,624.00){\usebox{\plotpoint}}
\multiput(473,624)(20.756,0.000){0}{\usebox{\plotpoint}}
\multiput(480,624)(20.756,0.000){0}{\usebox{\plotpoint}}
\put(489.82,624.00){\usebox{\plotpoint}}
\multiput(493,624)(20.756,0.000){0}{\usebox{\plotpoint}}
\multiput(500,624)(20.756,0.000){0}{\usebox{\plotpoint}}
\put(510.58,624.00){\usebox{\plotpoint}}
\multiput(513,624)(20.756,0.000){0}{\usebox{\plotpoint}}
\multiput(520,624)(20.756,0.000){0}{\usebox{\plotpoint}}
\put(531.33,624.00){\usebox{\plotpoint}}
\multiput(533,624)(20.756,0.000){0}{\usebox{\plotpoint}}
\multiput(540,624)(20.756,0.000){0}{\usebox{\plotpoint}}
\put(552.09,624.00){\usebox{\plotpoint}}
\multiput(553,624)(20.756,0.000){0}{\usebox{\plotpoint}}
\multiput(560,624)(20.756,0.000){0}{\usebox{\plotpoint}}
\put(572.84,624.00){\usebox{\plotpoint}}
\multiput(573,624)(20.756,0.000){0}{\usebox{\plotpoint}}
\multiput(580,624)(20.756,0.000){0}{\usebox{\plotpoint}}
\multiput(587,624)(20.756,0.000){0}{\usebox{\plotpoint}}
\put(593.60,624.00){\usebox{\plotpoint}}
\multiput(600,624)(20.756,0.000){0}{\usebox{\plotpoint}}
\multiput(607,624)(20.756,0.000){0}{\usebox{\plotpoint}}
\put(614.35,624.00){\usebox{\plotpoint}}
\multiput(620,624)(20.756,0.000){0}{\usebox{\plotpoint}}
\multiput(627,624)(20.756,0.000){0}{\usebox{\plotpoint}}
\put(635.11,624.00){\usebox{\plotpoint}}
\multiput(640,624)(20.756,0.000){0}{\usebox{\plotpoint}}
\multiput(647,624)(20.756,0.000){0}{\usebox{\plotpoint}}
\put(655.87,624.00){\usebox{\plotpoint}}
\multiput(660,624)(20.756,0.000){0}{\usebox{\plotpoint}}
\multiput(667,624)(20.756,0.000){0}{\usebox{\plotpoint}}
\put(676.62,624.00){\usebox{\plotpoint}}
\multiput(680,624)(20.756,0.000){0}{\usebox{\plotpoint}}
\multiput(687,624)(20.756,0.000){0}{\usebox{\plotpoint}}
\put(697.38,624.00){\usebox{\plotpoint}}
\multiput(700,624)(20.756,0.000){0}{\usebox{\plotpoint}}
\multiput(707,624)(20.756,0.000){0}{\usebox{\plotpoint}}
\put(718.13,624.00){\usebox{\plotpoint}}
\multiput(720,624)(20.756,0.000){0}{\usebox{\plotpoint}}
\multiput(727,624)(20.756,0.000){0}{\usebox{\plotpoint}}
\put(738.89,624.00){\usebox{\plotpoint}}
\multiput(740,624)(20.756,0.000){0}{\usebox{\plotpoint}}
\multiput(747,624)(20.756,0.000){0}{\usebox{\plotpoint}}
\put(759.64,624.00){\usebox{\plotpoint}}
\multiput(760,624)(20.756,0.000){0}{\usebox{\plotpoint}}
\multiput(767,624)(20.756,0.000){0}{\usebox{\plotpoint}}
\multiput(773,624)(20.756,0.000){0}{\usebox{\plotpoint}}
\put(780.40,624.00){\usebox{\plotpoint}}
\multiput(787,624)(20.756,0.000){0}{\usebox{\plotpoint}}
\multiput(793,624)(20.756,0.000){0}{\usebox{\plotpoint}}
\put(801.15,624.00){\usebox{\plotpoint}}
\multiput(807,624)(20.756,0.000){0}{\usebox{\plotpoint}}
\multiput(813,624)(20.756,0.000){0}{\usebox{\plotpoint}}
\put(821.91,624.00){\usebox{\plotpoint}}
\multiput(827,624)(20.756,0.000){0}{\usebox{\plotpoint}}
\multiput(833,624)(20.756,0.000){0}{\usebox{\plotpoint}}
\put(842.67,624.00){\usebox{\plotpoint}}
\multiput(847,624)(20.756,0.000){0}{\usebox{\plotpoint}}
\multiput(853,624)(20.756,0.000){0}{\usebox{\plotpoint}}
\put(863.42,624.00){\usebox{\plotpoint}}
\multiput(867,624)(20.756,0.000){0}{\usebox{\plotpoint}}
\multiput(873,624)(20.756,0.000){0}{\usebox{\plotpoint}}
\put(880,624){\usebox{\plotpoint}}
\end{picture}
\gnuplotcaptionskip
\caption{{\protect \captionfont $c({\bf p})$ at various momenta for
the D234(${2\over 3}$) action on a 2:1 (pion $\bullet$, rho $\circ$)
and the SW action on a 1:1 lattice (pion $\diamond$, rho $\times$).
For both actions $m_\rho/m_\pi \approx 1.47$ and $a_s \approx 0.32\, {\rm fm}$.
}}
\label{fig:c}
\end{figure}
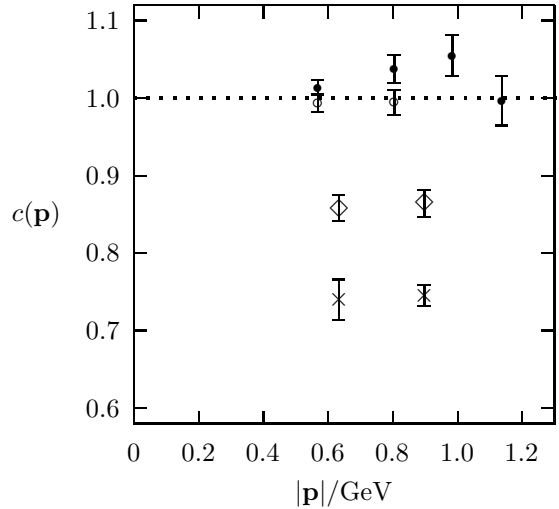
}

\figc

The anisotropic Symanzik-improved gluon action as well as the TI prescription
we used is described in~\cite{MorPea}. In addition to
$O(a_s^4)$ and $O(a_s^2 \alpha)$ errors, our gluon action has
$\O(a_t^2)$ errors from leaving out terms that give rise to
unphysical branches. For anisotropies $a_s/a_t \geq 2$ the $\O(a_t^2)$
errors seem to be small. However, we have noticed
that, as the coupling is varied, the ratio of string-tension to charmonium
determinations of $a_s$ shows variations of about 5\%.
This may be due to small violations of rotational
invariance, and we are in the process of non-perturbatively tuning the
gluon action to further reduce them~\cite{aniso}.

For our preliminary investigation of quenched QCD  on anisotropic
lattices, we have chosen to study the D234(${2\o 3}$) action 
(with the same TI prescription as for the glue)
on 2:1 lattices for light quarks, and on 3:1 lattices for heavy masses.
We use the SW action in various comparisons; both our as well as the results
of~\cite{SCRI} use TI.
For our general simulation methodology
we refer to~\cite{oldD234}.

For light quarks
we performed simulations at three couplings,
$\beta \! = \! 2.0, 2.15, 2.3$,
corresponding to lattice spacings $a_s= 0.41, 0.36, 0.32$ fm, if one uses
charmonium to set the scale \cite{aniso}.
The spatial size of our lattices is about 2.4~fm
(smaller for large quark masses).

The first question is whether there are large $\O(a^0)$ effects
violating space-time symmetry. As one can see in figure~\ref{fig:c},
after TI such effects are very
small.  
In addition, figures \ref{fig:c}
and \ref{fig:ccharm} show that D234(${2\o 3}$)
is also more continuum-like at momenta up to 1~GeV, indicating
that with TI the $\O(a^2)$ terms have roughly the right coefficients.

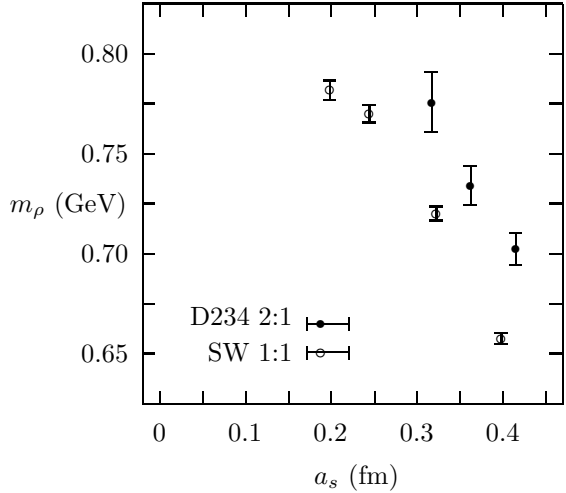
\begin{figure}[htbp]
\vskip -3mm
\setlength{\unitlength}{0.240900pt}
\ifx\plotpoint\undefined\newsavebox{\plotpoint}\fi
\begin{picture}(944,765)(0,0)
\font\gnuplot=cmr10 at 10pt
\gnuplot
\sbox{\plotpoint}{\rule[-0.200pt]{0.400pt}{0.400pt}}%
\put(220.0,113.0){\rule[-0.200pt]{4.818pt}{0.400pt}}
\put(860.0,113.0){\rule[-0.200pt]{4.818pt}{0.400pt}}
\put(220.0,192.0){\rule[-0.200pt]{4.818pt}{0.400pt}}
\put(198,192){\makebox(0,0)[r]{0.65}}
\put(860.0,192.0){\rule[-0.200pt]{4.818pt}{0.400pt}}
\put(220.0,270.0){\rule[-0.200pt]{4.818pt}{0.400pt}}
\put(860.0,270.0){\rule[-0.200pt]{4.818pt}{0.400pt}}
\put(220.0,349.0){\rule[-0.200pt]{4.818pt}{0.400pt}}
\put(198,349){\makebox(0,0)[r]{0.70}}
\put(860.0,349.0){\rule[-0.200pt]{4.818pt}{0.400pt}}
\put(220.0,428.0){\rule[-0.200pt]{4.818pt}{0.400pt}}
\put(860.0,428.0){\rule[-0.200pt]{4.818pt}{0.400pt}}
\put(220.0,506.0){\rule[-0.200pt]{4.818pt}{0.400pt}}
\put(198,506){\makebox(0,0)[r]{0.75}}
\put(860.0,506.0){\rule[-0.200pt]{4.818pt}{0.400pt}}
\put(220.0,585.0){\rule[-0.200pt]{4.818pt}{0.400pt}}
\put(860.0,585.0){\rule[-0.200pt]{4.818pt}{0.400pt}}
\put(220.0,663.0){\rule[-0.200pt]{4.818pt}{0.400pt}}
\put(198,663){\makebox(0,0)[r]{0.80}}
\put(860.0,663.0){\rule[-0.200pt]{4.818pt}{0.400pt}}
\put(220.0,742.0){\rule[-0.200pt]{4.818pt}{0.400pt}}
\put(860.0,742.0){\rule[-0.200pt]{4.818pt}{0.400pt}}
\put(247.0,113.0){\rule[-0.200pt]{0.400pt}{4.818pt}}
\put(247,68){\makebox(0,0){0}}
\put(247.0,722.0){\rule[-0.200pt]{0.400pt}{4.818pt}}
\put(314.0,113.0){\rule[-0.200pt]{0.400pt}{4.818pt}}
\put(314.0,722.0){\rule[-0.200pt]{0.400pt}{4.818pt}}
\put(382.0,113.0){\rule[-0.200pt]{0.400pt}{4.818pt}}
\put(382,68){\makebox(0,0){0.1}}
\put(382.0,722.0){\rule[-0.200pt]{0.400pt}{4.818pt}}
\put(449.0,113.0){\rule[-0.200pt]{0.400pt}{4.818pt}}
\put(449.0,722.0){\rule[-0.200pt]{0.400pt}{4.818pt}}
\put(516.0,113.0){\rule[-0.200pt]{0.400pt}{4.818pt}}
\put(516,68){\makebox(0,0){0.2}}
\put(516.0,722.0){\rule[-0.200pt]{0.400pt}{4.818pt}}
\put(584.0,113.0){\rule[-0.200pt]{0.400pt}{4.818pt}}
\put(584.0,722.0){\rule[-0.200pt]{0.400pt}{4.818pt}}
\put(651.0,113.0){\rule[-0.200pt]{0.400pt}{4.818pt}}
\put(651,68){\makebox(0,0){0.3}}
\put(651.0,722.0){\rule[-0.200pt]{0.400pt}{4.818pt}}
\put(718.0,113.0){\rule[-0.200pt]{0.400pt}{4.818pt}}
\put(718.0,722.0){\rule[-0.200pt]{0.400pt}{4.818pt}}
\put(786.0,113.0){\rule[-0.200pt]{0.400pt}{4.818pt}}
\put(786,68){\makebox(0,0){0.4}}
\put(786.0,722.0){\rule[-0.200pt]{0.400pt}{4.818pt}}
\put(853.0,113.0){\rule[-0.200pt]{0.400pt}{4.818pt}}
\put(853.0,722.0){\rule[-0.200pt]{0.400pt}{4.818pt}}
\put(220.0,113.0){\rule[-0.200pt]{158.994pt}{0.400pt}}
\put(880.0,113.0){\rule[-0.200pt]{0.400pt}{151.526pt}}
\put(220.0,742.0){\rule[-0.200pt]{158.994pt}{0.400pt}}
\put(111,427){\makebox(0,0){\lower 3mm\hbox{{$m_\rho$ (GeV) }}}}
\put(550,23){\makebox(0,0){\lower 7mm\hbox{{ $a_s ~({\rm fm})$}}}}
\put(220.0,113.0){\rule[-0.200pt]{0.400pt}{151.526pt}}
\put(456,239){\makebox(0,0)[r]{\raise 2mm\hbox{{D234 2:1}}}}
\put(500,239){\circle*{12}}
\put(806,356){\circle*{12}}
\put(735,456){\circle*{12}}
\put(674,587){\circle*{12}}
\put(478.0,239.0){\rule[-0.200pt]{15.899pt}{0.400pt}}
\put(478.0,229.0){\rule[-0.200pt]{0.400pt}{4.818pt}}
\put(544.0,229.0){\rule[-0.200pt]{0.400pt}{4.818pt}}
\put(806.0,331.0){\rule[-0.200pt]{0.400pt}{12.286pt}}
\put(796.0,331.0){\rule[-0.200pt]{4.818pt}{0.400pt}}
\put(796.0,382.0){\rule[-0.200pt]{4.818pt}{0.400pt}}
\put(735.0,425.0){\rule[-0.200pt]{0.400pt}{14.936pt}}
\put(725.0,425.0){\rule[-0.200pt]{4.818pt}{0.400pt}}
\put(725.0,487.0){\rule[-0.200pt]{4.818pt}{0.400pt}}
\put(674.0,540.0){\rule[-0.200pt]{0.400pt}{22.645pt}}
\put(664.0,540.0){\rule[-0.200pt]{4.818pt}{0.400pt}}
\put(664.0,634.0){\rule[-0.200pt]{4.818pt}{0.400pt}}
\put(456,194){\makebox(0,0)[r]{\raise 0mm\hbox{SW 1:1}}}
\put(500,194){\circle{12}}
\put(783,215){\circle{12}}
\put(681,412){\circle{12}}
\put(576,569){\circle{12}}
\put(514,606){\circle{12}}
\put(478.0,194.0){\rule[-0.200pt]{15.899pt}{0.400pt}}
\put(478.0,184.0){\rule[-0.200pt]{0.400pt}{4.818pt}}
\put(544.0,184.0){\rule[-0.200pt]{0.400pt}{4.818pt}}
\put(783.0,207.0){\rule[-0.200pt]{0.400pt}{4.095pt}}
\put(773.0,207.0){\rule[-0.200pt]{4.818pt}{0.400pt}}
\put(773.0,224.0){\rule[-0.200pt]{4.818pt}{0.400pt}}
\put(681.0,401.0){\rule[-0.200pt]{0.400pt}{5.300pt}}
\put(671.0,401.0){\rule[-0.200pt]{4.818pt}{0.400pt}}
\put(671.0,423.0){\rule[-0.200pt]{4.818pt}{0.400pt}}
\put(576.0,555.0){\rule[-0.200pt]{0.400pt}{6.745pt}}
\put(566.0,555.0){\rule[-0.200pt]{4.818pt}{0.400pt}}
\put(566.0,583.0){\rule[-0.200pt]{4.818pt}{0.400pt}}
\put(514.0,591.0){\rule[-0.200pt]{0.400pt}{7.227pt}}
\put(504.0,591.0){\rule[-0.200pt]{4.818pt}{0.400pt}}
\put(504.0,621.0){\rule[-0.200pt]{4.818pt}{0.400pt}}
\end{picture}
\gnuplotcaptionskip
\caption{{\protect \captionfont
The rho mass for the 2:1 D234(${2\o 3}$) and 1:1 SW~\protect\cite{SCRI}
actions.
}}
\label{fig:GeVrho}
\end{figure}

In figure~\ref{fig:GeVrho} we see that the D234(${2\o 3}$) rho mass, converted
to physical units
using the charmonium S$-$P splitting, 
is closer to the continuum value on coarse lattices than for SW.

The value of $J$~\cite{J} in table~\ref{tab:Jetc} is around $0.39(1)$ already
on the coarsest lattices (as for isotropic D234~\cite{oldD234});
SW reaches this value only for much finer lattices~\cite{SCRI}).
The $m_N/m_\rho$ ratio appears rather high, especially at $\beta=2.3$,
where it is larger than that of~\cite{oldD234} as well as~\cite{SCRI}.
However, this might be due to chiral fitting uncertainties
and problems with the glue.

\begin{table}[bt]
\setlength{\tabcolsep}{2mm}
\catcode`?=\active \def?{\kern\digitwidth}
\caption{{\protect\captionfont Comparison of 2:1 D234(${2\o 3}$)
and 1:1 SW~\protect\cite{SCRI} }}
\label{tab:Jetc}
\begin{tabular*}{75mm}{lllll}
\hline
\st Action & $\beta$ & $a$(fm) & ~~$J$  & $m_N/m_\rho$ \\
\hline
D234 & 2.0  &  0.41  &  0.380(5)    & 1.37(2) \\
D234 & 2.15 & 0.36   &  0.396(9)    & 1.32(5) \\
D234 & 2.3  & 0.32   &  0.375(6)(6) & 1.38(2)(3) \\
\hline
SW   & 6.8  & 0.40    &  0.345(4) & 1.457(15) \\
SW   & 7.75  & 0.18  &   0.386(9) & 1.31(3) \\
\hline
\end{tabular*}
\end{table}

\hide{
\begin{table}[bt]
\setlength{\tabcolsep}{2mm}
\catcode`?=\active \def?{\kern\digitwidth}
\caption{{\protect\captionfont Agreement of 2:1 and 3:1 D234(${2\o 3}$),
indicating that unphysical brances have little effect.
$R = a(2:1)/a(3:1) = 0.9894$ from the ratio of string tensions.
}}
\label{tab:ghosts}
\begin{tabular*}{75mm}{llll}
\hline
\st 
\st Particle & $ma(2:1)$ & $ R ma(3:1)$ & Matched $\pi$ \\
\hline
 P  &   1.528(3)   &   1.5078(45) &  1.5280(45)  \\
 V  &   1.942(4)   &   1.935(15)  & 1.951(15) \\
 A  &   3.02(4)    &   3.007(51)  & 3.023(51) \\
 N  &   3.000(14)  &   3.013(17)  & 3.034(17) \\
 D  &   3.26(2)    &   3.235(28)  & 3.257(28) \\
\hline
\end{tabular*}
\end{table}
}

\def\figccharm{
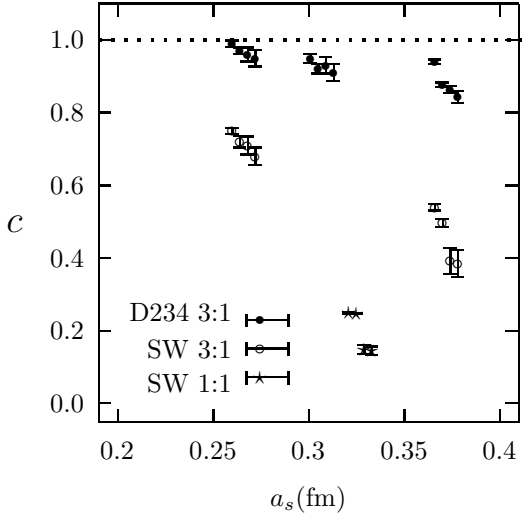
\begin{figure}[tbhp]
\vskip -8mm       
\setlength{\unitlength}{0.240900pt}
\ifx\plotpoint\undefined\newsavebox{\plotpoint}\fi
\sbox{\plotpoint}{\rule[-0.200pt]{0.400pt}{0.400pt}}%
\begin{picture}(944,793)(0,0)
\font\gnuplot=cmr10 at 10pt
\gnuplot
\sbox{\plotpoint}{\rule[-0.200pt]{0.400pt}{0.400pt}}%
\put(220.0,142.0){\rule[-0.200pt]{4.818pt}{0.400pt}}
\put(198,142){\makebox(0,0)[r]{$0.0$}}
\put(860.0,142.0){\rule[-0.200pt]{4.818pt}{0.400pt}}
\put(220.0,256.0){\rule[-0.200pt]{4.818pt}{0.400pt}}
\put(198,256){\makebox(0,0)[r]{$0.2$}}
\put(860.0,256.0){\rule[-0.200pt]{4.818pt}{0.400pt}}
\put(220.0,370.0){\rule[-0.200pt]{4.818pt}{0.400pt}}
\put(198,370){\makebox(0,0)[r]{$0.4$}}
\put(860.0,370.0){\rule[-0.200pt]{4.818pt}{0.400pt}}
\put(220.0,484.0){\rule[-0.200pt]{4.818pt}{0.400pt}}
\put(198,484){\makebox(0,0)[r]{$0.6$}}
\put(860.0,484.0){\rule[-0.200pt]{4.818pt}{0.400pt}}
\put(220.0,599.0){\rule[-0.200pt]{4.818pt}{0.400pt}}
\put(198,599){\makebox(0,0)[r]{$0.8$}}
\put(860.0,599.0){\rule[-0.200pt]{4.818pt}{0.400pt}}
\put(220.0,713.0){\rule[-0.200pt]{4.818pt}{0.400pt}}
\put(198,713){\makebox(0,0)[r]{$1.0$}}
\put(860.0,713.0){\rule[-0.200pt]{4.818pt}{0.400pt}}
\put(250.0,113.0){\rule[-0.200pt]{0.400pt}{4.818pt}}
\put(250,68){\makebox(0,0){$0.2$}}
\put(250.0,750.0){\rule[-0.200pt]{0.400pt}{4.818pt}}
\put(400.0,113.0){\rule[-0.200pt]{0.400pt}{4.818pt}}
\put(400,68){\makebox(0,0){$0.25$}}
\put(400.0,750.0){\rule[-0.200pt]{0.400pt}{4.818pt}}
\put(550.0,113.0){\rule[-0.200pt]{0.400pt}{4.818pt}}
\put(550,68){\makebox(0,0){$0.3$}}
\put(550.0,750.0){\rule[-0.200pt]{0.400pt}{4.818pt}}
\put(700.0,113.0){\rule[-0.200pt]{0.400pt}{4.818pt}}
\put(700,68){\makebox(0,0){$0.35$}}
\put(700.0,750.0){\rule[-0.200pt]{0.400pt}{4.818pt}}
\put(850.0,113.0){\rule[-0.200pt]{0.400pt}{4.818pt}}
\put(850,68){\makebox(0,0){$0.4$}}
\put(850.0,750.0){\rule[-0.200pt]{0.400pt}{4.818pt}}
\put(220.0,113.0){\rule[-0.200pt]{158.994pt}{0.400pt}}
\put(880.0,113.0){\rule[-0.200pt]{0.400pt}{158.271pt}}
\put(220.0,770.0){\rule[-0.200pt]{158.994pt}{0.400pt}}
\put(89,441){\makebox(0,0){\lower 5mm\hbox{{\Large {$c$}}}}}
\put(550,23){\makebox(0,0){\lower 8mm\hbox{{$a_s ({\rm fm})$}}}}
\put(220.0,113.0){\rule[-0.200pt]{0.400pt}{158.271pt}}
\put(430,273){\makebox(0,0)[r]{\raise 2mm\hbox{D234 3:1}}}
\put(474,273){\circle*{12}}
\put(430,707){\circle*{12}}
\put(442,696){\circle*{12}}
\put(454,690){\circle*{12}}
\put(466,684){\circle*{12}}
\put(452.0,273.0){\rule[-0.200pt]{15.899pt}{0.400pt}}
\put(452.0,263.0){\rule[-0.200pt]{0.400pt}{4.818pt}}
\put(518.0,263.0){\rule[-0.200pt]{0.400pt}{4.818pt}}
\put(430.0,702.0){\rule[-0.200pt]{0.400pt}{2.409pt}}
\put(420.0,702.0){\rule[-0.200pt]{4.818pt}{0.400pt}}
\put(420.0,712.0){\rule[-0.200pt]{4.818pt}{0.400pt}}
\put(442.0,691.0){\rule[-0.200pt]{0.400pt}{2.409pt}}
\put(432.0,691.0){\rule[-0.200pt]{4.818pt}{0.400pt}}
\put(432.0,701.0){\rule[-0.200pt]{4.818pt}{0.400pt}}
\put(454.0,679.0){\rule[-0.200pt]{0.400pt}{5.300pt}}
\put(444.0,679.0){\rule[-0.200pt]{4.818pt}{0.400pt}}
\put(444.0,701.0){\rule[-0.200pt]{4.818pt}{0.400pt}}
\put(466.0,671.0){\rule[-0.200pt]{0.400pt}{6.263pt}}
\put(456.0,671.0){\rule[-0.200pt]{4.818pt}{0.400pt}}
\put(456.0,697.0){\rule[-0.200pt]{4.818pt}{0.400pt}}
\put(553,684){\circle*{12}}
\put(565,667){\circle*{12}}
\put(577,673){\circle*{12}}
\put(589,661){\circle*{12}}
\put(553.0,677.0){\rule[-0.200pt]{0.400pt}{3.373pt}}
\put(543.0,677.0){\rule[-0.200pt]{4.818pt}{0.400pt}}
\put(543.0,691.0){\rule[-0.200pt]{4.818pt}{0.400pt}}
\put(565.0,660.0){\rule[-0.200pt]{0.400pt}{3.613pt}}
\put(555.0,660.0){\rule[-0.200pt]{4.818pt}{0.400pt}}
\put(555.0,675.0){\rule[-0.200pt]{4.818pt}{0.400pt}}
\put(577.0,660.0){\rule[-0.200pt]{0.400pt}{6.263pt}}
\put(567.0,660.0){\rule[-0.200pt]{4.818pt}{0.400pt}}
\put(567.0,686.0){\rule[-0.200pt]{4.818pt}{0.400pt}}
\put(589.0,648.0){\rule[-0.200pt]{0.400pt}{6.504pt}}
\put(579.0,648.0){\rule[-0.200pt]{4.818pt}{0.400pt}}
\put(579.0,675.0){\rule[-0.200pt]{4.818pt}{0.400pt}}
\put(748,679){\circle*{12}}
\put(760,643){\circle*{12}}
\put(772,635){\circle*{12}}
\put(784,624){\circle*{12}}
\put(748.0,675.0){\rule[-0.200pt]{0.400pt}{1.686pt}}
\put(738.0,675.0){\rule[-0.200pt]{4.818pt}{0.400pt}}
\put(738.0,682.0){\rule[-0.200pt]{4.818pt}{0.400pt}}
\put(760.0,639.0){\rule[-0.200pt]{0.400pt}{1.686pt}}
\put(750.0,639.0){\rule[-0.200pt]{4.818pt}{0.400pt}}
\put(750.0,646.0){\rule[-0.200pt]{4.818pt}{0.400pt}}
\put(772.0,629.0){\rule[-0.200pt]{0.400pt}{2.891pt}}
\put(762.0,629.0){\rule[-0.200pt]{4.818pt}{0.400pt}}
\put(762.0,641.0){\rule[-0.200pt]{4.818pt}{0.400pt}}
\put(784.0,614.0){\rule[-0.200pt]{0.400pt}{4.577pt}}
\put(774.0,614.0){\rule[-0.200pt]{4.818pt}{0.400pt}}
\put(774.0,633.0){\rule[-0.200pt]{4.818pt}{0.400pt}}
\put(430,228){\makebox(0,0)[r]{\lower 1mm\hbox{SW 3:1}}}
\put(474,228){\circle{12}}
\put(430,570){\circle{12}}
\put(442,553){\circle{12}}
\put(454,547){\circle{12}}
\put(466,530){\circle{12}}
\put(452.0,228.0){\rule[-0.200pt]{15.899pt}{0.400pt}}
\put(452.0,218.0){\rule[-0.200pt]{0.400pt}{4.818pt}}
\put(518.0,218.0){\rule[-0.200pt]{0.400pt}{4.818pt}}
\put(430.0,565.0){\rule[-0.200pt]{0.400pt}{2.409pt}}
\put(420.0,565.0){\rule[-0.200pt]{4.818pt}{0.400pt}}
\put(420.0,575.0){\rule[-0.200pt]{4.818pt}{0.400pt}}
\put(442.0,544.0){\rule[-0.200pt]{0.400pt}{4.336pt}}
\put(432.0,544.0){\rule[-0.200pt]{4.818pt}{0.400pt}}
\put(432.0,562.0){\rule[-0.200pt]{4.818pt}{0.400pt}}
\put(454.0,533.0){\rule[-0.200pt]{0.400pt}{6.745pt}}
\put(444.0,533.0){\rule[-0.200pt]{4.818pt}{0.400pt}}
\put(444.0,561.0){\rule[-0.200pt]{4.818pt}{0.400pt}}
\put(466.0,516.0){\rule[-0.200pt]{0.400pt}{6.745pt}}
\put(456.0,516.0){\rule[-0.200pt]{4.818pt}{0.400pt}}
\put(456.0,544.0){\rule[-0.200pt]{4.818pt}{0.400pt}}
\put(748,450){\circle{12}}
\put(760,426){\circle{12}}
\put(772,366){\circle{12}}
\put(784,362){\circle{12}}
\put(748.0,445.0){\rule[-0.200pt]{0.400pt}{2.409pt}}
\put(738.0,445.0){\rule[-0.200pt]{4.818pt}{0.400pt}}
\put(738.0,455.0){\rule[-0.200pt]{4.818pt}{0.400pt}}
\put(760.0,419.0){\rule[-0.200pt]{0.400pt}{3.132pt}}
\put(750.0,419.0){\rule[-0.200pt]{4.818pt}{0.400pt}}
\put(750.0,432.0){\rule[-0.200pt]{4.818pt}{0.400pt}}
\put(772.0,345.0){\rule[-0.200pt]{0.400pt}{9.877pt}}
\put(762.0,345.0){\rule[-0.200pt]{4.818pt}{0.400pt}}
\put(762.0,386.0){\rule[-0.200pt]{4.818pt}{0.400pt}}
\put(784.0,340.0){\rule[-0.200pt]{0.400pt}{10.359pt}}
\put(774.0,340.0){\rule[-0.200pt]{4.818pt}{0.400pt}}
\put(774.0,383.0){\rule[-0.200pt]{4.818pt}{0.400pt}}
\put(430,183){\makebox(0,0)[r]{\lower 4mm\hbox{SW 1:1}}}
\put(474,183){\makebox(0,0){$\star$}}
\put(613,284){\makebox(0,0){$\star$}}
\put(625,283){\makebox(0,0){$\star$}}
\put(637,226){\makebox(0,0){$\star$}}
\put(649,224){\makebox(0,0){$\star$}}
\put(452.0,183.0){\rule[-0.200pt]{15.899pt}{0.400pt}}
\put(452.0,173.0){\rule[-0.200pt]{0.400pt}{4.818pt}}
\put(518.0,173.0){\rule[-0.200pt]{0.400pt}{4.818pt}}
\put(613.0,283.0){\rule[-0.200pt]{0.400pt}{0.482pt}}
\put(603.0,283.0){\rule[-0.200pt]{4.818pt}{0.400pt}}
\put(603.0,285.0){\rule[-0.200pt]{4.818pt}{0.400pt}}
\put(625.0,282.0){\rule[-0.200pt]{0.400pt}{0.482pt}}
\put(615.0,282.0){\rule[-0.200pt]{4.818pt}{0.400pt}}
\put(615.0,284.0){\rule[-0.200pt]{4.818pt}{0.400pt}}
\put(637.0,219.0){\rule[-0.200pt]{0.400pt}{3.373pt}}
\put(627.0,219.0){\rule[-0.200pt]{4.818pt}{0.400pt}}
\put(627.0,233.0){\rule[-0.200pt]{4.818pt}{0.400pt}}
\put(649.0,218.0){\rule[-0.200pt]{0.400pt}{3.132pt}}
\put(639.0,218.0){\rule[-0.200pt]{4.818pt}{0.400pt}}
\put(639.0,231.0){\rule[-0.200pt]{4.818pt}{0.400pt}}
\sbox{\plotpoint}{\rule[-0.500pt]{1.000pt}{1.000pt}}%
\put(220,713){\usebox{\plotpoint}}
\put(220.00,713.00){\usebox{\plotpoint}}
\multiput(227,713)(20.756,0.000){0}{\usebox{\plotpoint}}
\multiput(233,713)(20.756,0.000){0}{\usebox{\plotpoint}}
\put(240.76,713.00){\usebox{\plotpoint}}
\multiput(247,713)(20.756,0.000){0}{\usebox{\plotpoint}}
\multiput(253,713)(20.756,0.000){0}{\usebox{\plotpoint}}
\put(261.51,713.00){\usebox{\plotpoint}}
\multiput(267,713)(20.756,0.000){0}{\usebox{\plotpoint}}
\multiput(273,713)(20.756,0.000){0}{\usebox{\plotpoint}}
\put(282.27,713.00){\usebox{\plotpoint}}
\multiput(287,713)(20.756,0.000){0}{\usebox{\plotpoint}}
\multiput(293,713)(20.756,0.000){0}{\usebox{\plotpoint}}
\put(303.02,713.00){\usebox{\plotpoint}}
\multiput(307,713)(20.756,0.000){0}{\usebox{\plotpoint}}
\multiput(313,713)(20.756,0.000){0}{\usebox{\plotpoint}}
\put(323.78,713.00){\usebox{\plotpoint}}
\multiput(327,713)(20.756,0.000){0}{\usebox{\plotpoint}}
\multiput(333,713)(20.756,0.000){0}{\usebox{\plotpoint}}
\put(344.53,713.00){\usebox{\plotpoint}}
\multiput(347,713)(20.756,0.000){0}{\usebox{\plotpoint}}
\multiput(353,713)(20.756,0.000){0}{\usebox{\plotpoint}}
\put(365.29,713.00){\usebox{\plotpoint}}
\multiput(367,713)(20.756,0.000){0}{\usebox{\plotpoint}}
\multiput(373,713)(20.756,0.000){0}{\usebox{\plotpoint}}
\put(386.04,713.00){\usebox{\plotpoint}}
\multiput(387,713)(20.756,0.000){0}{\usebox{\plotpoint}}
\multiput(393,713)(20.756,0.000){0}{\usebox{\plotpoint}}
\put(406.80,713.00){\usebox{\plotpoint}}
\multiput(407,713)(20.756,0.000){0}{\usebox{\plotpoint}}
\multiput(413,713)(20.756,0.000){0}{\usebox{\plotpoint}}
\multiput(420,713)(20.756,0.000){0}{\usebox{\plotpoint}}
\put(427.56,713.00){\usebox{\plotpoint}}
\multiput(433,713)(20.756,0.000){0}{\usebox{\plotpoint}}
\multiput(440,713)(20.756,0.000){0}{\usebox{\plotpoint}}
\put(448.31,713.00){\usebox{\plotpoint}}
\multiput(453,713)(20.756,0.000){0}{\usebox{\plotpoint}}
\multiput(460,713)(20.756,0.000){0}{\usebox{\plotpoint}}
\put(469.07,713.00){\usebox{\plotpoint}}
\multiput(473,713)(20.756,0.000){0}{\usebox{\plotpoint}}
\multiput(480,713)(20.756,0.000){0}{\usebox{\plotpoint}}
\put(489.82,713.00){\usebox{\plotpoint}}
\multiput(493,713)(20.756,0.000){0}{\usebox{\plotpoint}}
\multiput(500,713)(20.756,0.000){0}{\usebox{\plotpoint}}
\put(510.58,713.00){\usebox{\plotpoint}}
\multiput(513,713)(20.756,0.000){0}{\usebox{\plotpoint}}
\multiput(520,713)(20.756,0.000){0}{\usebox{\plotpoint}}
\put(531.33,713.00){\usebox{\plotpoint}}
\multiput(533,713)(20.756,0.000){0}{\usebox{\plotpoint}}
\multiput(540,713)(20.756,0.000){0}{\usebox{\plotpoint}}
\put(552.09,713.00){\usebox{\plotpoint}}
\multiput(553,713)(20.756,0.000){0}{\usebox{\plotpoint}}
\multiput(560,713)(20.756,0.000){0}{\usebox{\plotpoint}}
\put(572.84,713.00){\usebox{\plotpoint}}
\multiput(573,713)(20.756,0.000){0}{\usebox{\plotpoint}}
\multiput(580,713)(20.756,0.000){0}{\usebox{\plotpoint}}
\multiput(587,713)(20.756,0.000){0}{\usebox{\plotpoint}}
\put(593.60,713.00){\usebox{\plotpoint}}
\multiput(600,713)(20.756,0.000){0}{\usebox{\plotpoint}}
\multiput(607,713)(20.756,0.000){0}{\usebox{\plotpoint}}
\put(614.35,713.00){\usebox{\plotpoint}}
\multiput(620,713)(20.756,0.000){0}{\usebox{\plotpoint}}
\multiput(627,713)(20.756,0.000){0}{\usebox{\plotpoint}}
\put(635.11,713.00){\usebox{\plotpoint}}
\multiput(640,713)(20.756,0.000){0}{\usebox{\plotpoint}}
\multiput(647,713)(20.756,0.000){0}{\usebox{\plotpoint}}
\put(655.87,713.00){\usebox{\plotpoint}}
\multiput(660,713)(20.756,0.000){0}{\usebox{\plotpoint}}
\multiput(667,713)(20.756,0.000){0}{\usebox{\plotpoint}}
\put(676.62,713.00){\usebox{\plotpoint}}
\multiput(680,713)(20.756,0.000){0}{\usebox{\plotpoint}}
\multiput(687,713)(20.756,0.000){0}{\usebox{\plotpoint}}
\put(697.38,713.00){\usebox{\plotpoint}}
\multiput(700,713)(20.756,0.000){0}{\usebox{\plotpoint}}
\multiput(707,713)(20.756,0.000){0}{\usebox{\plotpoint}}
\put(718.13,713.00){\usebox{\plotpoint}}
\multiput(720,713)(20.756,0.000){0}{\usebox{\plotpoint}}
\multiput(727,713)(20.756,0.000){0}{\usebox{\plotpoint}}
\put(738.89,713.00){\usebox{\plotpoint}}
\multiput(740,713)(20.756,0.000){0}{\usebox{\plotpoint}}
\multiput(747,713)(20.756,0.000){0}{\usebox{\plotpoint}}
\put(759.64,713.00){\usebox{\plotpoint}}
\multiput(760,713)(20.756,0.000){0}{\usebox{\plotpoint}}
\multiput(767,713)(20.756,0.000){0}{\usebox{\plotpoint}}
\multiput(773,713)(20.756,0.000){0}{\usebox{\plotpoint}}
\put(780.40,713.00){\usebox{\plotpoint}}
\multiput(787,713)(20.756,0.000){0}{\usebox{\plotpoint}}
\multiput(793,713)(20.756,0.000){0}{\usebox{\plotpoint}}
\put(801.15,713.00){\usebox{\plotpoint}}
\multiput(807,713)(20.756,0.000){0}{\usebox{\plotpoint}}
\multiput(813,713)(20.756,0.000){0}{\usebox{\plotpoint}}
\put(821.91,713.00){\usebox{\plotpoint}}
\multiput(827,713)(20.756,0.000){0}{\usebox{\plotpoint}}
\multiput(833,713)(20.756,0.000){0}{\usebox{\plotpoint}}
\put(842.67,713.00){\usebox{\plotpoint}}
\multiput(847,713)(20.756,0.000){0}{\usebox{\plotpoint}}
\multiput(853,713)(20.756,0.000){0}{\usebox{\plotpoint}}
\put(863.42,713.00){\usebox{\plotpoint}}
\multiput(867,713)(20.756,0.000){0}{\usebox{\plotpoint}}
\multiput(873,713)(20.756,0.000){0}{\usebox{\plotpoint}}
\put(880,713){\usebox{\plotpoint}}
\end{picture}
\gnuplotcaptionskip
\caption{{\protect \captionfont
Effective velocity of light $c({\bf p})$ for the D234(${2\o 3}$) and SW
actions on various lattices at the charm mass, $|{\bf p}| \approx 670\,$MeV.
Each quadruplet represents, from left to right,
the $\eta_c$ (``pion''), $J/\psi$ (``rho''),
``nucleon'' and ``delta''.
For clarity we have slightly displaced the x-coordinates
around the actual $a_s$. 
}}
\label{fig:ccharm}
\end{figure}
}

\def\figHFS{
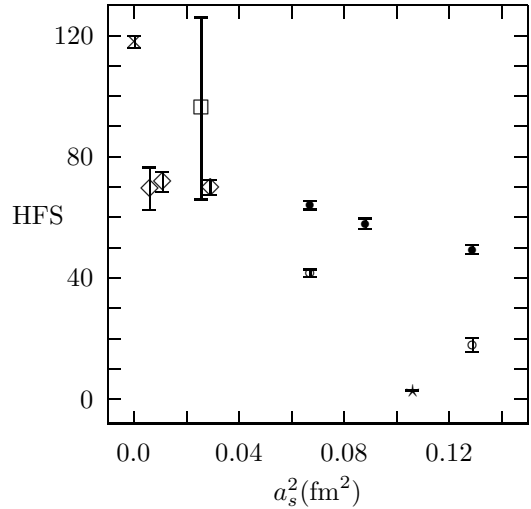
\begin{figure}[htbp]
\setlength{\unitlength}{0.240900pt}
\ifx\plotpoint\undefined\newsavebox{\plotpoint}\fi
\begin{picture}(944,793)(0,0)
\font\gnuplot=cmr10 at 10pt
\gnuplot
\sbox{\plotpoint}{\rule[-0.200pt]{0.400pt}{0.400pt}}%
\put(220.0,151.0){\rule[-0.200pt]{4.818pt}{0.400pt}}
\put(198,151){\makebox(0,0)[r]{0}}
\put(860.0,151.0){\rule[-0.200pt]{4.818pt}{0.400pt}}
\put(220.0,199.0){\rule[-0.200pt]{4.818pt}{0.400pt}}
\put(860.0,199.0){\rule[-0.200pt]{4.818pt}{0.400pt}}
\put(220.0,246.0){\rule[-0.200pt]{4.818pt}{0.400pt}}
\put(860.0,246.0){\rule[-0.200pt]{4.818pt}{0.400pt}}
\put(220.0,294.0){\rule[-0.200pt]{4.818pt}{0.400pt}}
\put(860.0,294.0){\rule[-0.200pt]{4.818pt}{0.400pt}}
\put(220.0,342.0){\rule[-0.200pt]{4.818pt}{0.400pt}}
\put(198,342){\makebox(0,0)[r]{40}}
\put(860.0,342.0){\rule[-0.200pt]{4.818pt}{0.400pt}}
\put(220.0,389.0){\rule[-0.200pt]{4.818pt}{0.400pt}}
\put(860.0,389.0){\rule[-0.200pt]{4.818pt}{0.400pt}}
\put(220.0,437.0){\rule[-0.200pt]{4.818pt}{0.400pt}}
\put(860.0,437.0){\rule[-0.200pt]{4.818pt}{0.400pt}}
\put(220.0,484.0){\rule[-0.200pt]{4.818pt}{0.400pt}}
\put(860.0,484.0){\rule[-0.200pt]{4.818pt}{0.400pt}}
\put(220.0,532.0){\rule[-0.200pt]{4.818pt}{0.400pt}}
\put(198,532){\makebox(0,0)[r]{80}}
\put(860.0,532.0){\rule[-0.200pt]{4.818pt}{0.400pt}}
\put(220.0,580.0){\rule[-0.200pt]{4.818pt}{0.400pt}}
\put(860.0,580.0){\rule[-0.200pt]{4.818pt}{0.400pt}}
\put(220.0,627.0){\rule[-0.200pt]{4.818pt}{0.400pt}}
\put(860.0,627.0){\rule[-0.200pt]{4.818pt}{0.400pt}}
\put(220.0,675.0){\rule[-0.200pt]{4.818pt}{0.400pt}}
\put(860.0,675.0){\rule[-0.200pt]{4.818pt}{0.400pt}}
\put(220.0,722.0){\rule[-0.200pt]{4.818pt}{0.400pt}}
\put(198,722){\makebox(0,0)[r]{120}}
\put(860.0,722.0){\rule[-0.200pt]{4.818pt}{0.400pt}}
\put(261.0,113.0){\rule[-0.200pt]{0.400pt}{4.818pt}}
\put(261,68){\makebox(0,0){0.0}}
\put(261.0,750.0){\rule[-0.200pt]{0.400pt}{4.818pt}}
\put(344.0,113.0){\rule[-0.200pt]{0.400pt}{4.818pt}}
\put(344.0,750.0){\rule[-0.200pt]{0.400pt}{4.818pt}}
\put(426.0,113.0){\rule[-0.200pt]{0.400pt}{4.818pt}}
\put(426,68){\makebox(0,0){0.04}}
\put(426.0,750.0){\rule[-0.200pt]{0.400pt}{4.818pt}}
\put(509.0,113.0){\rule[-0.200pt]{0.400pt}{4.818pt}}
\put(509.0,750.0){\rule[-0.200pt]{0.400pt}{4.818pt}}
\put(591.0,113.0){\rule[-0.200pt]{0.400pt}{4.818pt}}
\put(591,68){\makebox(0,0){0.08}}
\put(591.0,750.0){\rule[-0.200pt]{0.400pt}{4.818pt}}
\put(674.0,113.0){\rule[-0.200pt]{0.400pt}{4.818pt}}
\put(674.0,750.0){\rule[-0.200pt]{0.400pt}{4.818pt}}
\put(756.0,113.0){\rule[-0.200pt]{0.400pt}{4.818pt}}
\put(756,68){\makebox(0,0){0.12}}
\put(756.0,750.0){\rule[-0.200pt]{0.400pt}{4.818pt}}
\put(839.0,113.0){\rule[-0.200pt]{0.400pt}{4.818pt}}
\put(839.0,750.0){\rule[-0.200pt]{0.400pt}{4.818pt}}
\put(220.0,113.0){\rule[-0.200pt]{158.994pt}{0.400pt}}
\put(880.0,113.0){\rule[-0.200pt]{0.400pt}{158.271pt}}
\put(220.0,770.0){\rule[-0.200pt]{158.994pt}{0.400pt}}
\put(111,441){\makebox(0,0){\raise 0mm\hbox{HFS}}}
\put(550,23){\makebox(0,0){\lower 6mm\hbox{$a^2_s ({\rm fm}^2)$}}}
\put(220.0,113.0){\rule[-0.200pt]{0.400pt}{158.271pt}}
\put(538,456){\circle*{12}}
\put(625,427){\circle*{12}}
\put(793,386){\circle*{12}}
\put(538.0,449.0){\rule[-0.200pt]{0.400pt}{3.132pt}}
\put(528.0,449.0){\rule[-0.200pt]{4.818pt}{0.400pt}}
\put(528.0,462.0){\rule[-0.200pt]{4.818pt}{0.400pt}}
\put(625.0,419.0){\rule[-0.200pt]{0.400pt}{3.854pt}}
\put(615.0,419.0){\rule[-0.200pt]{4.818pt}{0.400pt}}
\put(615.0,435.0){\rule[-0.200pt]{4.818pt}{0.400pt}}
\put(793.0,379.0){\rule[-0.200pt]{0.400pt}{3.373pt}}
\put(783.0,379.0){\rule[-0.200pt]{4.818pt}{0.400pt}}
\put(783.0,393.0){\rule[-0.200pt]{4.818pt}{0.400pt}}
\put(538,349){\circle{12}}
\put(793,236){\circle{12}}
\put(538.0,343.0){\rule[-0.200pt]{0.400pt}{2.891pt}}
\put(528.0,343.0){\rule[-0.200pt]{4.818pt}{0.400pt}}
\put(528.0,355.0){\rule[-0.200pt]{4.818pt}{0.400pt}}
\put(793.0,225.0){\rule[-0.200pt]{0.400pt}{5.300pt}}
\put(783.0,225.0){\rule[-0.200pt]{4.818pt}{0.400pt}}
\put(783.0,247.0){\rule[-0.200pt]{4.818pt}{0.400pt}}
\put(699,164){\makebox(0,0){$\star$}}
\put(699.0,164.0){\usebox{\plotpoint}}
\put(689.0,164.0){\rule[-0.200pt]{4.818pt}{0.400pt}}
\put(689.0,165.0){\rule[-0.200pt]{4.818pt}{0.400pt}}
\put(286,481){\raisebox{-.8pt}{\makebox(0,0){$\Diamond$}}}
\put(306,492){\raisebox{-.8pt}{\makebox(0,0){$\Diamond$}}}
\put(381,483){\raisebox{-.8pt}{\makebox(0,0){$\Diamond$}}}
\put(286.0,448.0){\rule[-0.200pt]{0.400pt}{16.140pt}}
\put(276.0,448.0){\rule[-0.200pt]{4.818pt}{0.400pt}}
\put(276.0,515.0){\rule[-0.200pt]{4.818pt}{0.400pt}}
\put(306.0,477.0){\rule[-0.200pt]{0.400pt}{7.468pt}}
\put(296.0,477.0){\rule[-0.200pt]{4.818pt}{0.400pt}}
\put(296.0,508.0){\rule[-0.200pt]{4.818pt}{0.400pt}}
\put(381.0,472.0){\rule[-0.200pt]{0.400pt}{5.541pt}}
\put(371.0,472.0){\rule[-0.200pt]{4.818pt}{0.400pt}}
\put(371.0,495.0){\rule[-0.200pt]{4.818pt}{0.400pt}}
\put(367,608){\raisebox{-.8pt}{\makebox(0,0){$\Box$}}}
\put(367.0,465.0){\rule[-0.200pt]{0.400pt}{68.897pt}}
\put(357.0,465.0){\rule[-0.200pt]{4.818pt}{0.400pt}}
\put(357.0,751.0){\rule[-0.200pt]{4.818pt}{0.400pt}}
\put(262,713){\makebox(0,0){$\times$}}
\put(262.0,703.0){\rule[-0.200pt]{0.400pt}{4.577pt}}
\put(252.0,703.0){\rule[-0.200pt]{4.818pt}{0.400pt}}
\put(252.0,722.0){\rule[-0.200pt]{4.818pt}{0.400pt}}
\end{picture}
\vspace{-1mm}
\gnuplotcaptionskip
\caption{{\protect \captionfont
Quenched charmonium hyper-fine splitting (in MeV):
D234(${2\o 3}$) 3:1 ($\bullet$), SW 3:1 ($\circ$),
SW 1:1 ($\star$),
FNAL ($\diamond$)~\protect\cite{FNAL}, NRQCD ($\Box$)~\protect\cite{NRQCD},
experiment ($\times$).
}}
\label{fig:HFS}
\end{figure}
}

\figccharm
\figHFS

Dramatic improvements are seen for the charmonium spectrum on anisotropic
lattices. In figures~\ref{fig:ccharm} and~\ref{fig:HFS} we show results for
D234(${2\o 3}$)
on 3:1 lattices and compare them with SW on 3:1 and 1:1 lattices. As expected,
isotropic SW shows $O(1)$ scaling violations; SW on 3:1 is much better, and
D234(${2\o 3}$) on 3:1 is much better still.

Overall, anisotropic lattices are clearly well suited to coarse-lattice
studies of heavy quarks and glueballs~\cite{MorPea}.
We have also seen some improvement in the rho mass for light quarks.
However, our present results have systematic uncertainties in the glue and
scale setting, so that it remains to be seen how big such improvements
will ultimately be.
For improved quark actions to be accurate on
 isotropic or anisotropic coarse lattices
it might be necessary to eliminate $\O(a)$
quantum errors by non-perturbative tuning of the coefficients.

\section{Non-perturbative Tuning and the $r$-Test}

Our simulations so far rely on TI to estimate the coefficients in
an action. We are now starting work on non-perturbative tuning.
For an anisotropic action such a tuning might be necessary already at
$\O(a^0)$, since the spatial and temporal first order derivatives might
renormalize differently. As figure~\ref{fig:c} shows, after TI
this effect is small; the remainder could   easily     be tuned by
introducing a ``bare velocity of light'' into the action and 
demanding
$c({\bf p})\seq 1$ for small masses and momenta.

At $\O(a)$ one has to tune the clover term.
Let us for the moment restrict
ourselves to the isotropic case, where there are only two terms at $\O(a)$,
which we write as
$-{a r \over 2}( \sum_\mu \Delta_\mu + {\om\over 2}\, \si \! \cdot \! F)$.
\hide{
\beq 
-{a r \over 2} \,
\biggl( \sum_\mu \Delta_\mu + {\om\over 2}\, \si \! \cdot \! F \biggl) ~.
\eeq
}
The coefficient $r$, say, can be adjusted to any desired value by a field
transformation. To eliminate quantum errors at $\O(a)$ one therefore has 
only to tune the (relative) clover coefficient $\om$.
A tuning method was presented in~\cite{LPCAC} for the SW action on
Wilson glue, for which it was found that $\om$ 
(at zero quark mass)
is significantly larger
than the TI estimate. Note that if this were also the case for the D234 action,
it might solve the problem of its low rho mass.

An alternative method could be based on tuning $\om$ 
so that the spectrum of
the action is invariant under (small) changes of the Wilson $r$ parameter,
cf.~sect.~1. 
If $\om$ were independent of $r$ (as it is classically) this will only be
true for the correct non-perturbative $\om$. 
The         fact that $\om$ 
presumably depends on $r$ at the quantum level, complicates the conversion of 
the $r$-test from a rough consistency check into a full tuning method.
%
However, this is not a problem of principle, and we are presently exploring
this idea. 

On    anisotropic lattices there is the additional complication,
for both tuning and the $r$-test,
 that
the temporal and spatial parts of the $\O(a)$ terms
can suffer a relative renormalization. 
Assuming that TI makes this
renormalization small, as we have seen for the $\O(a^0)$ terms, we can
however
also apply the $r$-test to anisotropic lattices.
We have obtained preliminary results 
for the D234 actions of sect.~3.  On a 2:1
lattice with $a_s \seq 0.36$~fm we measured the difference between the
rho masses for $r\seq {2\o 3}$ and $r\seq 1$ 
for (i) $\om=0$, and (ii) the tree-level tadpole improved $\om$.
Tuning $m_\rho/m_\pi$ to be $1.3$ in all cases,
we find      
the change in the rho mass (in lattice units) to be $0.104(10)$
in the first case, and $0.012(4)$ in the second.
This suggests that the correct 
$\om$      
is
somewhat, but
not dramatically larger than the TI estimate
(even taking into
account other results indicating that $\om$ increases with decreasing mass).

\end{document}